\documentclass[5p,final,authoryear,twocolumn,dvipsnames]{elsarticle}
\pdfoutput=1
\usepackage[utf8]{inputenc}
\usepackage{graphicx}
\usepackage{amssymb}
\usepackage{algorithm}
\usepackage{algpseudocode}

\usepackage{array}
\usepackage{amsmath}
\usepackage{todonotes}
\usepackage{hyperref}
\usepackage{makecell}
\usepackage{url}
\usepackage{multirow}
\usepackage{annotate-equations}
\usepackage{booktabs}
\usepackage{caption}
\usepackage{subcaption}
\usepackage{fancyref}
\usepackage{svg}
\usepackage{multicol}
\usepackage{adjustbox}

\newcommand{\classifiermodel}[1]{f_{\theta}(#1)}
\newcommand{\imagevaluemodel}[1]{f_\zeta(#1)}
\newcommand{\image}[1]{\mathit{I}_{#1}}

\journal{Medical Image Analysis}

\begin{document}

\begin{frontmatter}


\title{MINT: A wrapper to make multi-modal and multi-image AI models interactive}


\makeatletter
\def\@author#1{\g@addto@macro\elsauthors{\normalsize%
    \def\baselinestretch{1}%
    \upshape\authorsep#1\unskip\textsuperscript{%
      \ifx\@fnmark\@empty\else\unskip\sep\@fnmark\let\sep=,\fi
      \ifx\@corref\@empty\else\unskip\sep\@corref\let\sep=,\fi
      }%
    \def\authorsep{\unskip,\space}%
    \global\let\@fnmark\@empty
    \global\let\@corref\@empty  
    \global\let\sep\@empty}%
    \@eadauthor={#1}
}
\makeatother

\author[]{Jan~Freyberg$^1$\corref{cor2}}
\author[]{Abhijit~Guha~Roy$^1$\corref{cor2}}
\author[]{Terry~Spitz$^1$\corref{cor2}}
\author[]{Beverly~Freeman$^1$}
\author[]{Mike~Schaekermann$^1$}
\author[]{Patricia~Strachan$^1$}
\author[]{Eva~Schnider$^1$}
\author[]{Renee~Wong$^1$}
\author[]{Dale~R~Webster$^1$}
\author[]{Alan~Karthikesalingam$^1$}
\author[]{Yun~Liu$^1$}
\author[]{Krishnamurthy~Dvijotham$^2$\corref{cor1}}
\author[]{Umesh~Telang$^1$\corref{cor1}}
\cortext[cor2]{{J. Freyberg, A. Guha Roy and T. Spitz are joint first authors.}}
\cortext[cor1]{{U. Telang and K. Dvijotham are joint last authors. \\
Correspondence to: 
\{janfreyberg, agroy, terryspitz\}@google.com }}

\address{$^1$Google Research, Health AI, $^2$Google DeepMind}

\begin{abstract}

During the diagnostic process, doctors incorporate multimodal information including imaging and the medical history - and similarly medical AI development has increasingly become multimodal. In this paper we tackle a more subtle challenge: doctors take a targeted medical history to obtain only the most pertinent pieces of information; how do we enable AI to do the same? We develop a wrapper method named MINT (\textbf{M}ake your model \textbf{INT}eractive) that automatically determines what pieces of information are most valuable at each step, and ask for only the most useful information. 
We demonstrate the efficacy of MINT wrapping a skin disease prediction model, where multiple images and a set of optional answers to $25$ standard metadata questions (i.e., structured medical history) are used by a multi-modal deep network to provide a differential diagnosis.
We show that MINT can identify whether metadata inputs are needed and if so, which question to ask next. We also demonstrate that when collecting multiple images, MINT can identify if an additional image would be beneficial, and if so, which type of image to capture.
We showed that MINT reduces the number of metadata and image inputs needed by 82\% and 36.2\% respectively, while maintaining predictive performance. Using real-world AI dermatology system data, we show that needing fewer inputs can retain users that may otherwise fail to complete the system submission and drop off without a diagnosis.
Qualitative examples show MINT can closely mimic the step-by-step decision making process of a clinical workflow and how this is different for straight forward cases versus more difficult, ambiguous cases.
Finally we demonstrate how MINT is robust to different underlying multi-model classifiers and can be easily adapted to user requirements without significant model re-training.

\end{abstract}


\end{frontmatter}

\section{Introduction}
\label{sec:intro}

Deep learning for medical applications increasingly involves structured, tabular data to augment unstructured input data such as images and signals.
This reflects the nature of medical care, in which doctors tend to make decisions based on multi-modal information instead of purely based on imaging or physiological data. For example, alongside capturing images or scans, patient information such as demographics, patient history, non-visual symptoms, signs, and test results can be critical in the decision making process. Often, even the unstructured data comes in multiple views, such as frontal and lateral chest X-rays, or photographs of skin lesions from multiple perspectives or in multiple locations. Acquisition of the complete battery of information involves collection of extraneous information and its associated costs.

By contrast, clinicians in practice approach data collection during diagnosis as an iterative process. At every step they assimilate all information gathered so far to determine the top candidate diagnoses. They then ask specific questions or gather specific data points to rule in or rule out diagnoses. This process can be characterised by three attributes: (i) A \textbf{sequential decision making process} to gather additional information, (ii) A \textbf{personalised flow} -- which questions to ask in what order -- which varies across patient journeys, based on data acquired, and (iii) \textbf{Identifying when to stop} the acquisition process when a reliable diagnosis has been obtained. In this way, the diagnostic data collection process is targeted and enables both faster diagnosis and avoids superfluous testing.

In this paper, we focus on medical AI and the human-AI interaction during data acquisition and consider how active feature acquisition (AFA, also known as dynamic feature selection) can improve interactions with patients when multiple views and multiple modalities are required for prediction.

\begin{figure}[t]
\includegraphics[scale=0.38,page=1]{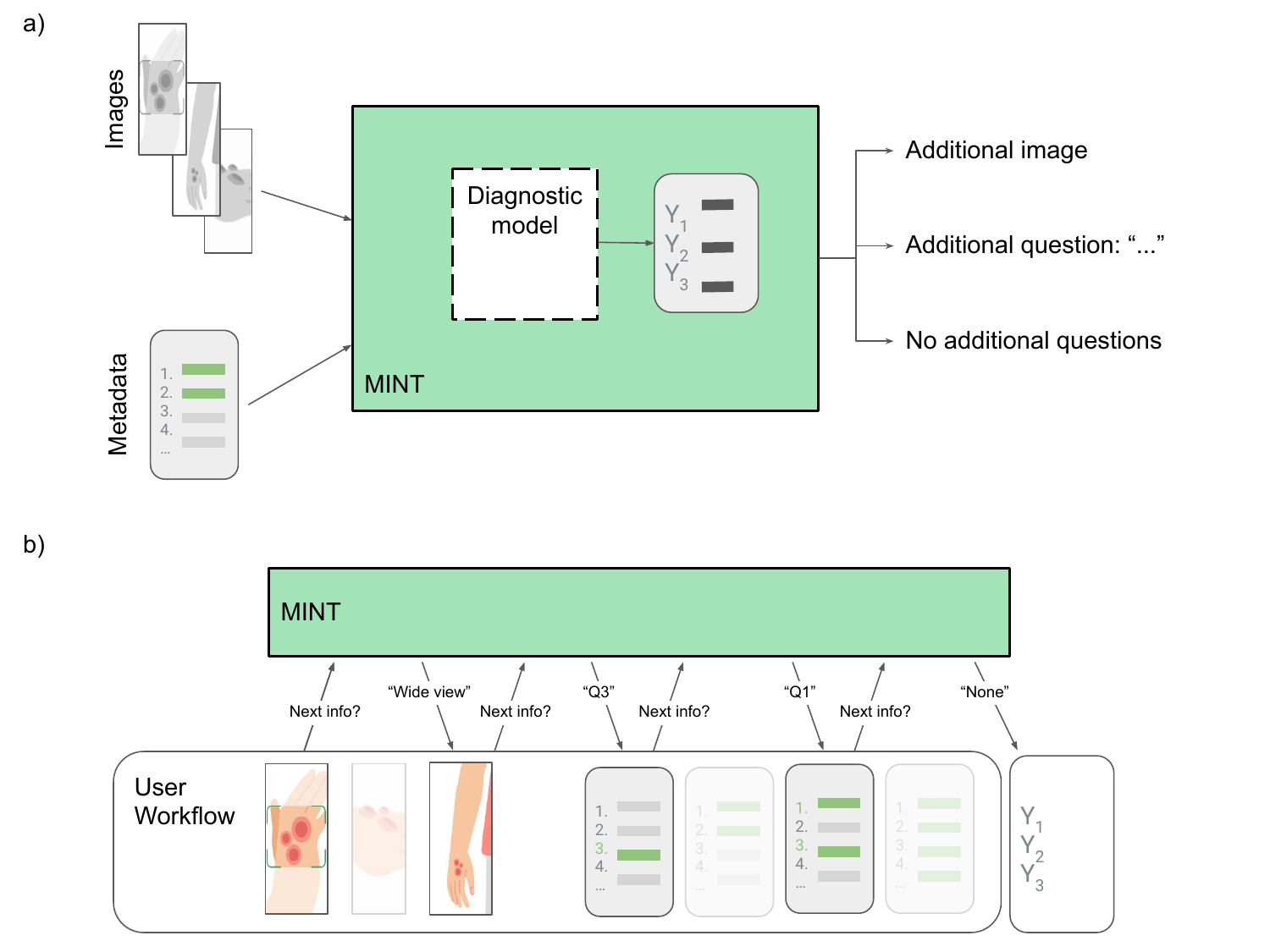}
\caption{\textbf{a)} The MINT framework wraps a multi-modal diagnostic model for interactive use, taking a partial set of images and metadata as inputs and returning the next image type, question or early stopping token. \textbf{b)} The workflow consists of interactive data gathering; in each round all data collected so far is passed to MINT. If MINT requests additional data, the next image or answer is collected from the user, otherwise the workflow stops and returns a final result from the diagnostic model.}
\end{figure}

\subsection{Related work}

We consider the field of AFA as most relevant to the work presented here. The goal of AFA is to acquire a subset of features, in a cost sensitive manner, in order to maximise test time performance. Most work estimates the information value of a feature at each step. This tends to be approached using either Reinforcement Learning (RL) or greedy approaches.

\textbf{RL approaches} have the benefit that agents can theoretically be trained to maximise the performance of the full sequential feature selection, rather than just that of the next feature. Previous work treats the diagnostic model as a black box reward evaluator \citep{bernardino2022reinforcement}, or by sharing representations and training the diagnostic and acquisition model jointly \citep{jointafa}. However, despite the promise of RL to optimise the sequential acquisition task as a whole, recent work shows that greedy approaches \citep{chauhan2022interactive, covert2023learning} and even static approaches \citep{erion2022coai, chauhan2022interactive} can outperform RL-based methods. Whether RL is competitive varies significantly across tasks, indicating that the methods are sensitive to the dimensionality and cardinality of the dataset.

One particularly relevant paper is \citep{kossen2022active}. The authors focus on multi-modal data, specifically temporal datasets, where data are acquired over time. This is relevant for medical scenarios, where over the course of a patient's treatment, different data may become important at different points. Multi-modal data are handled using the PerceiverIO architecture, and the authors demonstrate that their approach can structurally handle sequential data with missing features. However, the acquisition agent produces static behaviour rather than "personalised" behaviour for each example.

\textbf{Greedy approaches} instead maximise information gain for each selected feature one interaction at a time. For example, \citep{ma2018eddi} and \citep{zannone2019odin} model the information gain by predicting missing features using variational autoencoders (VAEs), and using the predicted features to calculate information gain. We recently proposed a particular formulation for models relying on intermediate concepts in which we use confidence and value of the concepts to determine their importance \citep{chauhan2022interactive}. Recently, Covert et al. \citep{covert2023learning} aim to use the advantages of greedy selection in training a reinforcement learning algorithm. They train a policy to greedily select the next best feature, improving the stability of the resulting policy and succeeding in applying it to a broad range of datasets.

The majority of existing work covers tabular data for historic and practical reasons. AFA was developed for web data in the e-commerce space (see e.g. \citealp{zheng2002active}), and before deep learning made high-dimensional data modelling more commonplace. Practically, many metrics and objectives optimised by many AFA algorithms are tractable only for low-dimensional data, such as mutual information between features and outcome (see e.g. \citealp{Chen2015b-pmlr}). Recent work in medical domains has involved multi-modal classification \citep{akrout2018improving, akrout2019improving}, but only tabular data was acquired interactively.

Work that involves actively acquiring image data often focuses on single-image, pixel-wise data acquisition (e.g. \citealp{li2021towards, ghosh2023difa, li2020dynamic}). This is an approximation of active perception \citep{bajcsy2018revisiting}, a problem of real-world perception by robot agents that may not be able to perceive a whole scene all at once, but is has limited relevance in many medical domain tasks.

Most relevant to our work is recent work by Akrout and colleagues (\citeyear{akrout2018improving, akrout2019improving}). The authors focus on dermatology classification using images. Using single-image diagnostic data, the authors train a classification model that estimates the probabilities of each disease. They then use an independent dataset for identifying structured questions related to the diseases they are diagnosing, which they term the knowledge graph. The authors update the probabilities produced by the image model using the knowledge graph by multiplying the conditional probability of a disease given a structured piece of information with the probability estimated by the image model. This means they do not require the image model and the structured updating to be trained on the same dataset.

The authors then test two approaches towards actively acquiring these structured questions. First, as they update probabilities, they can calculate the information gain for each structured question directly \citep{akrout2018improving}. Second, in order to estimate the benefit of questions in a non-greedy fashion, they use RL to identify the next question to ask \citep{akrout2019improving}. The authors see remarkable success, with the RL model improving over the greedy model by 22\%. In contrast to this work, we focus on a multi-modal model in which structured questions are used as input to the deep learning classifier directly. Depending on the specific dataset used, this allows the diagnostic model to compare image features and structured information when making diagnostic predictions and enables better fusion of structured information.

\subsection{Our contributions}

In this article, we propose \textbf{MINT}: \emph{\textbf{M}ake your model \mbox{\textbf{Int}eractive}}, an interactive AI framework which can mimic the iterative information acquisition process used in a clinical scenario.
We focus on the application of dermatology condition classification (as described in \citealp{liu2020deep} and \citealp{roy2022does}). We use multiple smartphone captured photographs and structured medical history metadata as inputs with the aim of providing potential differential diagnosis conditions as output.
MINT acts as a wrapper which can operate on top of any multi-modal/ multi-view ML model and make it interactive. This makes our solution easy to use in a scalable manner, without retraining the underlying model.
This enables ML models to iteratively acquire the most relevant information from the user, and thus reduces inference time data acquisition cost caused by querying for irrelevant or redundant data.

In this article, we:
\begin{itemize}
    \item Extend active feature acquisition to the multi-view, multi-modal setting encountered in many medical applications.
    \item Produce a simple but effective approach to actively acquiring photographs of complex skin problems that can require views from different perspectives.
    \item Develop an algorithm for active tabular data acquisition conditioned on high-dimensional image data.
    \item Evaluate our approach using a real-world dermatology dataset and show that our approach increases the absolute number of users supplied with a correct diagnosis by reducing the user drop-off rate while maintaining a high accuracy.
\end{itemize}

\section{Method}
\label{sec:method}
\subsection{Problem Formulation}
\label{sec:problem}

Let us consider a multi-modal and multi-view dataset $\mathcal{D} = \{ (\mathcal{X}_1, y_1), \dots, (\mathcal{X}_N, y_N) \}$ with $N$ samples, where the input $\mathcal{X}_i = \{ \image{1}, \dots, \image{j}, m_1, \dots, m_k\}^i$ represents a set of multi-modal inputs consisting of $j$ images $\mathcal{I}_i = \{ \image{1}, \dots, \image{j} \}$ and $k$ tabular metadata $\mathcal{M}_i = \{ m_1, \dots, m_k\}$, and $y$ indicates the class label.
Let $\classifiermodel{ \cdot }$ be a multi-modal ML model, with weights $\theta$ which learns to predict class $y$ from an input set $\mathcal{X}$, i.e. $f_{\theta}:\mathcal{X} \rightarrow y$.

In this article, we propose the MINT framework. MINT wraps around a trained model $\classifiermodel{\cdot}$ and makes it interactive in nature.
The wrapping process involves introducing two changes to the model input pipeline. First, each input from $\mathcal{X}$ is acquired one at a time in a step-by-step order with MINT iteratively establishing its sequence. This change introduces a personalised input flow for each patient similar to clinical setups.
Second, at every input acquisition step MINT estimates the value of incorporating additional inputs and can terminate the acquisition process early if no additional performance gains are expected. This change can potentially reduce the number of inputs required significantly.

MINT involves learning a value estimator $f_{\text{MINT}}(\cdot)$ and a value threshold $T$. 
At any step of the input acquisition process, we use $f_{\text{MINT}}(\cdot)$ to estimate the value for all possible inputs yet to acquire. We then acquire the input with highest estimated value (i.e. the most informative input) at each step. If the highest estimated value is below threshold $T$, we stop the acquisition process.

We find parameters for the value estimator $f_{\text{MINT}}(\cdot)$ and the threshold $T$ by optimising for two competing objectives. The first objective $\mathcal{O}_1$ in Eq.~\ref{eq:objective-1} indicates the difference in average cardinality of the total possible inputs to the inputs actually acquired by MINT. This can be written as:

\vspace{2em}
\begin{equation}
\mathcal{O}_1: \mathbb{E}_i[\eqnmarkbox[Aquamarine]{Xi}{| \mathcal{X}_i |}] - \mathbb{E}_i[\eqnmarkbox[BurntOrange]{Xihat}{| \hat{\mathcal{X}}_i |}] 
\annotate[yshift=0.7em]{above,}{Xi}{maximum input size}
\annotate[yshift=-0.7em]{below,}{Xihat}{acquired input size}
\label{eq:objective-1}
\end{equation} 
\vspace{1.2em}

The second objective $\mathcal{O}_2$ in Eq.~\ref{eq:objective-2} indicates the expected change in model performance using all possible inputs versus using only inputs acquired by MINT. This can be written as:

\vspace{2em}
\begin{equation}
\mathcal{O}_2: \mathbb{E}_i[\|
    \eqnmarkbox[Aquamarine]{Lfull}{\mathcal{L}(f_{\theta}, \mathcal{X}_i, y_i)}
    - \eqnmarkbox[BurntOrange]{Lhat}{\mathcal{L}(f_{\theta}, \hat{\mathcal{X}}_i, y_i)}
    \|] 
\annotate[yshift=0.7em,]{above,left}{Lfull}{performance with\\max input}
\annotate[yshift=0.7em]{above,right}{Lhat}{performance with\\acquired input}
\label{eq:objective-2}
\end{equation}

\noindent where $\mathcal{L}(\cdot)$ indicates the function to compute the performance metric.

MINT can be calibrated based on two use-cases. Task~1 shown in Eq.~\ref{eq:task1} considers the situation where the user provides a tolerance $\epsilon_2$ on the permissible drop in model performance ($\mathcal{O}_2$). The acquired input size is then maximally reduced ($\mathcal{O}_1$) while keeping the performance drop below $\epsilon_2$. Task~2 shown in Eq.~\ref{eq:task2} considers a situation where the user specifies a required average reduction in acquired inputs $\epsilon_1$. MINT then minimises the reduction in model performance $\mathcal{O}_2$.

\vspace{2em}
\begin{align}
\text{Task 1: } \max (\mathcal{O}_1) \text{  with } \mathcal{O}_2 &\leq \eqnmarkbox[Mulberry]{Eps2}{\epsilon_2}
\annotate[yshift=0.7em,]{above,left}{Eps2}{permissible performance drop}
\label{eq:task1} \\
\text{Task 2: } \min (\mathcal{O}_2) \text{  with } \mathcal{O}_1 &\geq \eqnmarkbox[RoyalBlue]{Eps1}{\epsilon_1}
\annotate[yshift=-0.7em,]{below,left}{Eps1}{desired number of\\whithheld inputs}
\label{eq:task2}
\end{align}
\vspace{2em}

\subsection{Multi-modal classification model}
\label{subsec:multimodal_model}
In this article, we focus on the application of classification of dermatological conditions. 
We use a multi-view, multi-modal model for the task, conceptualised as $f_{\theta}:\mathcal{X} \rightarrow y$.
Such multi-modal models are commonly decomposed into an image embedding model ($g(\cdot)$), and a metadata embedding, fusion and classification model ($h(\cdot)$), such that $\classifiermodel{\mathcal{X}_i} = h(g(\mathcal{I}_i), \mathcal{M}_i)$.
While this is the most common type of model in use by medical AI practitioners, this architecture is not a requirement for MINT to work. More details about the model training process can be found in Appendix ~\ref{app:model_training}. Details of the model are provided below.

\noindent
\textbf{Multi-image embedding: }
We use a wide ResNet-BiT $101 \times 3$ pre-trained on ImageNet-21k~\citep{kolesnikov2020big}, as the feature extractor $g(\cdot)$.
For all images for a case $\{ \image{1}, \dots, \image{j}\}$, we estimate an image embedding $\{ \mathbf{v}_1, \dots, \mathbf{v}_j\}$ where each $\mathbf{v}_{j} \in \mathbb{R}^{6044}$. We average-pool the available image embeddings for each case to get a case-level image embedding: $\hat{\mathbf{v}} = \mathbb{E}[\{\mathbf{v}_1, \dots, \mathbf{v}_j \}]$, where $\hat{\mathbf{v}} \in \mathbb{R}^{6044}$. This average-pooling ensures the model is agnostic to the number of input images.

\noindent
\textbf{Structured metadata embedding: }
In medical classifier models metadata is often collected as an input vector alongside images. These data consist of either scalar inputs (patient age) represented as $\mathbb{R}^1$, or categorical inputs in a multidimensional binary representation as $\mathbb{B}^{c}$, where $c$ is the number of possible answers for a given multi-categorical input, including an option of \emph{unknown}. For example, a question such as `Do you have fever?', with answers \emph{Yes/ No/ Unknown} is a $\mathbb{B}^{3}$ vector, while a question about Fitzpatrick skin type is a $\mathbb{B}^{7}$ vector (scale of 1-6 with an unknown option). For patient age, we use the median age in our training data as a place-holder for \emph{Unknown}.

In this article, answers to the 25 metadata questions (including age) in $\mathcal{M}$ are encoded in this way and concatenated, resulting in a vector $\mathbf{m} \in  \mathbb{R}^{1}~+~\mathbb{B}^{100}$.

\noindent
\textbf{Multi-modal fusion and Classification head: }
In this part, we first combine the case-level embeddings of images $\hat{\mathbf{v}}$ and metadata $\mathbf{m}$ using a fusion strategy, then pass them to a classifier head.
We compare two fusion methods as choices for combining multi-modal information: feature-wise linear modulation (FiLM, see~\citealp{perez2018film}) and concatenation with case-level embedding. Our proposed framework is agnostic to the choice of fusion strategy.

The classifier head is a two-layer MLP with hidden size $1024$ and output size $288$ (number of classes), which after softmax yields the class probabilities.

\subsection{Interactive input process using MINT}
\label{sec:mint_method}

MINT (Make your model INTeractive) is a light-weight scalable wrapper around any pre-trained multi-modal classifier to make its input acquisition process interactive.
In Algorithm~\ref{algo:mint_process}, we show the interactive input process for MINT.

Across both domains of input, we apply a simple strategy: to estimate the prospective value of a new piece of data, and to decide whether to actually acquire that piece of information based on its value estimate and a threshold (see Algorithm~\ref{algo:mint_estimate_value}).

\begin{algorithm}
\caption{MINT Interactive input process}
\begin{algorithmic}[1]
\State $\mathcal{X} \gets$ Permissible set of inputs \Comment{\textit{user specified}}
\State $\mathcal{\hat{X}} \gets \emptyset$ \Comment{\textit{input acquired by MINT}}
\State $\classifiermodel{\cdot} \gets$ classifier model
\State $d \gets 0$ \Comment{value of next action}
\State $\mathcal{\hat{X}} \gets \mathcal{\hat{X}} \cup \image{1}$ \Comment{\textit{start interaction; (random) first image}}
\While{$d \geq 0 \; \text{and} \; |\mathcal{\hat{X}}| < |\mathcal{X}|$}
    \State $\mathbf{d} \gets [0]_{1\times |\mathcal{X} \backslash \mathcal{\hat{X}}|}$ \Comment{\textit{value vector}}
    \For{$x$ in $\mathcal{X} \backslash \mathcal{\hat{X}}$} \Comment{\textit{loop over remaining inputs}}
        \State $\mathbf{d}[x] \gets $ \textsc{EstimateValue}($\mathcal{\hat{X}}$, $x$, $\classifiermodel{\cdot}$)
    \EndFor
    \State $x_{\text{next}}, d \gets $ argmax$(\mathbf{d})$, $\text{max}(\mathbf{d})$
    \State $\mathcal{\hat{X}} \gets \mathcal{\hat{X}} \cup \textit{Answer}(x_{\text{next}})$ \Comment{\textit{next input acquired}}
\EndWhile
\State return $\classifiermodel{\mathcal{\hat{X}}}$ \Comment{\textit{final prediction}}
\end{algorithmic}
\label{algo:mint_process}
\end{algorithm}

In brief, we assign the predictive distribution change with a particular input information as its value, and try to estimate it. Depending on the type of input (image or metadata), we propose two different strategies to estimate the value of acquiring additional input. For details, see Sections \ref{subsec:int_image} and \ref{subsec:int_metadata}.

\begin{algorithm*}
\caption{EstimateValue function $f_{\text{MINT}}(\cdot)$}
\begin{multicols}{2}
\begin{algorithmic}[1]
\State $\mathcal{I} \gets \text{Permissible image types}$
\State $\mathcal{M}_{cat} \gets$ Permissible categorical metadata
\State $\mathcal{M}_{real} \gets$ Permissible real-valued metadata
\State $\mathcal{D} \gets$ Distance metric to estimate value.
\State $T_\mathcal{I} \gets$ threshold for image value
\State $T_\mathcal{M} \gets$ threshold for metadata value
\State $\classifiermodel{\cdot} \gets$ classifier model
\State $\imagevaluemodel{\cdot} \gets$ estimator for value of image types
\Procedure{EstimateValue}{\par
\hskip \algorithmicindent $\mathcal{\hat{X}}$: current inputs, \par
\hskip \algorithmicindent $x$: prospective input, \par
\hskip \algorithmicindent $\classifiermodel{\cdot}$: diagnostic model}
\State $p_{\text{current}} \gets \classifiermodel{\mathcal{\hat{X}}}$ \Comment{\textit{current predictions}}
\If{$x \in \mathcal{I}$} \Comment{\textit{images}}
\State return $\imagevaluemodel{\mathcal{\hat{X}}, p_{\text{current}}, x} - T_\mathcal{I}$\par
\hspace{1em} 
\ElsIf{$x \in \mathcal{M}_{cat} \cup \mathcal{M}_{real}$} \Comment{\textit{metadata}}
\State $\mathbf{v} \gets [ \cdot ]$ \Comment{\textit{value vector}}
\If{$x \in \mathcal{M}_{cat}$}
\For{possible values $x_n$ of $x$}
\State $p_{\text{new}} \gets \classifiermodel{\mathcal{\hat{X}} \cup x_n}$
\State $\text{Append } \mathcal{D}( p_{\text{current}}, p_{\text{new}}) \text{ to } \mathbf{v}$
\EndFor
\ElsIf{$x \in \mathcal{M}_{real}$}
\For{percentiles $P_{10\%}, P_{50\%}, P_{90\%}$ of $x$}
\State $p_{\text{new}} \gets \classifiermodel{\mathcal{\hat{X}} \cup P \text{ of } x}$
\State $\text{Append } \mathcal{D}( p_{\text{current}}, p_{\text{new}}) \text{ to } \mathbf{v}$
\EndFor
\EndIf
\State return $\mathbb{E}(\mathbf{v}) - T_\mathcal{M}$
\EndIf
\EndProcedure
\end{algorithmic}
\end{multicols}
\label{algo:mint_estimate_value}
\end{algorithm*}

MINT's algorithm \ref{algo:mint_estimate_value} has several hyperparameters, which we tune using a validation set. First, the metric $\mathcal{D}$ used to calculate the distance between $p_{\text{current}}$ and $p_{\text{new}}$. We compare (i) Kullback-Leibler (KL) divergence, (ii) Jensen Shannon (JS) distance, and (iii) Absolute difference in predictive entropy. Second are the thresholds for selecting additional images or metadata, $T_{\{\mathcal{I}, \mathcal{M}\}}$, which we set by first selecting appropriate reductions in performance $\eta_{\{1, 2\}}$ in Equations \ref{eq:task1} and \ref{eq:task2} and finding threshold values that satisfy the objectives. Lastly, a statistical model $\imagevaluemodel{\cdot}$ is required to estimate the value of different images. We describe the development of this model in the next section.

\subsubsection{Interactive image acquisition}
\label{subsec:int_image}

The vast majority of information for predicting dermatological conditions lies in the image(s) \citep{liu2020deep}. Thus, we start the interactive acquisition process using MINT with images. Note that this is not a requirement.
We pose the interactive image acquisition process as a 3-way decision making task: (i) input a near-shot (NS) image, (ii) input a far-shot (FS) image, and (iii) no more images required. 
We repeatedly decide to either ask the user for additional images (NS/ FS), or stop and continue to the next stage of the workflow. 
NS/ FS serve as instructions to the user when taking the images, such as whether a skin photograph should be taken from close-up or further away respectively.
We train a supervised statistical model $\imagevaluemodel{\cdot}$ to accomplish this task.
We pose the learning task as a regression task predicting the value of an additional image, given all past image acquired.
For a given case, let us assume the model has already provided $N$ images $\{ I_i, \dots, I_N \}$. At the $(N+1)^{th}$ interaction $\imagevaluemodel{\{ I_i, \dots, I_N \}, \text{NS/ FS}}$ will provide the value of acquiring the corresponding image. Based on the predicted value, the next recommended action is asked of the user, including not asking for further images if the value is below threshold $\mathcal{T}_{\mathcal{I}}$.

\textbf{Model Training: }
We train $\imagevaluemodel{\cdot}$ using the validation split (on which the base multi-modal model was not trained). We split this in two sets to $75\%$ train and $25\%$ tune sets.
For each case $\mathcal{X}_i$, every image $I_j$ is a associated with a view-label $I_{\text{view}}^j \in \mathbb{B}^2$, indicating if it's near-shot or far-shot.
For each case, every subset with size greater than one from the set of available images can serve as a data-point for model training. Each subset of size $(j+1)$ can serve as a data-point for $j^{th}$ interaction, i.e. $\{I_1, \dots, I_j \} \rightarrow I_{j+1}$. 
We use the classifier model to estimate the predictions $p_{1:j}$ and $p_{1:j+1}$ before and after the interaction. Using the ground-truth label $y$, we define the value of the $j^{th}$ interaction as the change in predictive distance from the label before and after the interaction, i.e.
\begin{equation}
    \text{Value}^{j \rightarrow j+1} = \mathcal{D}(p_{1:j}, y) - \mathcal{D}(p_{1:j+1}, y).
\end{equation}

As an input for this $j^{th}$ interaction $\mathbf{z}^{j}$ can be set as, 
\begin{equation}
    \mathbf{z}^{{j \rightarrow j+1}} = \text{concat}[\hat{\mathbf{v}}_{1:j}; p_{1:j}; I_{\text{view}}^j; I_{\text{view}}^{j+1}].
\end{equation}

For each case, we can generate multiple data-points corresponding to different types of transitions. We construct a dataset for a regression task by creating data-points for different cases in the validation set, which is used to train $\imagevaluemodel{\cdot}$. As a choice of model, we use a Random Forest regressor~\citep{breiman2001random} for our task. Note that this is a design choice and may vary for different applications.

\textbf{Model tuning and evaluation:}
We can tune the value threshold $\mathcal{T}_{\mathcal{I}}$ on the tune sub-set based on input preference (Task 1 or Task 2) as discussed in Sec.~\ref{sec:problem}.
During evaluation, at each stage of interaction, we make two decisions. Firstly, the model decides whether an additional image will improve model prediction. Secondly, if an additional image is needed, the model needs to provide instructions to the user on whether a near-shot (NS) or a far-shot (FS) image is needed.
For this we estimate the value of both a NS and a FS next image at every stage of the interaction as

\begin{align}
    \text{Value}^{j \rightarrow j+1}_{\text{NS}} &= \imagevaluemodel{\mathbf{z}^{{j \rightarrow j+1}}(I_{\text{view}}^{j+1} = NS)} \\
    \text{Value}^{j \rightarrow j+1}_{\text{FS}} &= \imagevaluemodel{\mathbf{z}^{{j \rightarrow j+1}}(I_{\text{view}}^{j+1} = FS)}
\end{align}

Using these values, we average them to estimate the value of any additional image
\begin{equation}
    \text{Value}^{j \rightarrow j+1} = \mathbb{E}[\{ \text{Value}^{j \rightarrow j+1}_{\text{NS}}, \text{Value}^{j \rightarrow j+1}_{\text{FS}} \}].
\end{equation}

Following this, we decide if another image is needed:

\begin{align}
    \text{Value}^{j \rightarrow j+1} &\leq \mathcal{T}_{\mathcal{I}} \rightarrow \text{Early stopping}, \\
    \text{Value}^{j \rightarrow j+1} &> \mathcal{T}_{\mathcal{I}} \rightarrow \text{Take another image}.
\end{align}

If another image is needed, we decide on its instruction/ view label as 
\begin{align}
    \text{Value}^{j \rightarrow j+1}_{\text{NS}} &\geq \text{Value}^{j \rightarrow j+1}_{\text{FS}} \rightarrow \text{Take NS image}, \\
    \text{Value}^{j \rightarrow j+1}_{\text{NS}} &< \text{Value}^{j \rightarrow j+1}_{\text{FS}} \rightarrow \text{Take FS image}.
\end{align}

Note that the start of the interactive process is by providing a random image.

We compare our model based approach to a model confidence based interactive baseline. As a measure of model confidence we use max of softmax (MSP) as a signal~\citep{hendrycks2016baseline}. We use the MSP of the current prediction $\text{MSP}(p_{1:j})$ as the value for next image $\text{Value}^{j \rightarrow j+1}$. We set $\mathcal{T}_{\mathcal{I}}$ on the tune set. We compare MINT against this confidence based approach in Sec.~\ref{subsec:int_img_result}.

Note that, for each case in our dataset, we have different numbers of available images and varying NS and FS images. For our simulation experiments we set some boundary conditions. For a case with available $j$ images, if MINT asked for another image at the  $j^{th}$ interaction, early stopping was triggered due to unavailability.
Also, at any stage of the interaction, if MINT asks for a FS/ NS image and no unused image with that view label is available, we randomly provide MINT with any remaining unused image.

\subsubsection{Interactive tabular metadata acquisition}
\label{subsec:int_metadata}

In Algorithm~\ref{algo:mint_process}, we show how our MINT framework can interactively figure out the best metadata question to acquire and update the model predictions in a step-by-step process.

At any stage of the metadata interactive process, we sweep over all unanswered metadata questions ($\mathcal{X} \: \backslash \: \mathcal{\hat{X}}$) and estimate the expected value each of them would provide to the underlying model. We select the metadata that provides the highest value ($x_{\text{next}}$). We then ask the user to provide the answer, and update the model predictions.

\label{subsec:value_estimation}

We describe the process of value estimation in Algorithm~\ref{algo:mint_estimate_value}. As metadata input is structured, all possible answers for a given question $x_n$ can be enumerated (i.e. $\{x_n^1, \dots x_n^d\}$). We can modify the metadata embedding by considering all possible answers and estimating predictive probabilities $\{ p_{x_n}^1, \dots, p_{x_n}^d \}$, where $p_{x_n}^1 = \classifiermodel{\hat{\mathcal{X}} \cup x_n^1}$.
 
We define the value of a metadata $t_i$ with an answer $t_i^1$ as $\mathcal{D}(p_{\text{current}}, p_i^1)$. We the value for the metadata $x_n$ as the expected value for all possible answers, i.e. 

\begin{equation}
    \text{Value}_{t_i} = \mathbb{E}[\{\mathcal{D}(p_{\text{current}}, p^1_{x_n}), \dots,  \mathcal{D}(p_{\text{current}}, p_{x_n}^d)\}]
\end{equation}

At the end of an interactive step, the output can be either the next most informative question to ask to the user or to inform the user that no more input is needed and show the final prediction (see Algorithm \ref{algo:mint_process}). We refer to this feature as \textit{early stopping}. Early stopping is triggered when the value of the next best question is below the threshold $T_\mathcal{M}$. We set the value of $T$ using the tune subset of the validation set, by optimising Equations \ref{eq:task1} or \ref{eq:task2}.

We compare our proposed interactive approach against two baselines:
\begin{itemize}
    \item Random baseline: In this baseline for each case we acquire the metadata answer in a random sequence and report the average performance metrics computed over multiple trials.
    \item Global sequence: We estimate a static sequence of metadata questions which optimises performance for the whole validation set, i.e. the best possible sequence of questions for the whole validation set on average.
\end{itemize}

\subsection{Real-world cost of acquiring information}

One of the key aims of MINT is to reduce user burden. \emph{User burden} is a complex concept. In medical care in particular, burden can stem from the user's feelings about their care more generally, the need to seek care in the first place, and the implications of a diagnosis. In this paper, we focus on the burden of asking unnecessary and possibly invasive questions during the process of developing a differential diagnosis.

Specifically, we focus the proportion of users who drop off the submission flow of a non-diagnostic dermatology AI system  \citep{dermassist_2022}, i.e. do not complete a submission. While a user may drop off the submission flow for a variety of reasons, we are motivated to mitigate drop-off by reducing the number of irrelevant questions.

The system requires users to submit 3 images, and asks users to complete all metadata questions modelled in \ref{subsec:int_metadata}. The user is free to skip any of these. The 25 metadata questions modelled are grouped into 6 separate screens.

Metrics from this system report how many users reach which screen, including the final \emph{Results} screen. We estimate the user burden of requesting an image, or asking a particular question, as the proportion of users who drop off the submission flow at the screen on which the information is requested.

In order to estimate the reduction in drop-off rates using MINT, we simulate cases in our evaluation set as passing through the submission flow. If an image or piece of metadata $x$ is requested by MINT, we randomly drop cases from the flow with probability $p^{drop}_x$, i.e. the proportion of cases that drop off on the screen this question appears on, as estimated using the system's metrics. Note that as we only have these metrics at a screen level, if multiple questions that appear on the same screen are asked, we apply $p^{drop}_x$ only once. We do this simulation 1000 times, and compare the proportion of cases estimated to be dropped using MINT with the proportion of cases estimated to be dropped if all information is requested.

This is an imperfect estimate of the true reduction in user drop-off, as it makes several assumptions: \textbf{(1)} that $p^{drop}_X$ is independent, \textbf{(2)} that the sequencing of metadata questions has no impact, and \textbf{(3)} that the $p^{drop}$ for images is equally distributed across the three images requested by the system, which only reports metrics for image submission as a whole.

\section{Data description and experimental setup}
\label{sec:data}

In this article, we use the de-identified dataset used in~\citep{liu2020deep} for model development. Cases in this dataset were collected from $17$ different sites across California and Hawaii.
The dataset consists of $17\,454$ cases, with the ground truth generated by aggregating the diagnoses from multiple board certified dermatologists. The detailed annotation process is presented in~\citep{liu2020deep}.
Note that for simplicity, we removed the cases that had multiple conditions or that had multiple primary diagnoses in the ground truth.
The underlying model was trained on $12\,335$ cases. 
We use $3\,020$ cases as validation set and report performance on a test set with $2\,099$ cases.
We maintain a patient-level separation between the training, validation and test sets.
Each case in the dataset consists of multiple RGB images. The images were captured by either medical assistants or nurse practitioners using consumer-grade digital cameras.
The number of images captured and the type of images (near-shot/ far-shot etc) per case were decided by the clinician.
The images exhibit a large amount of variation in terms of affected anatomic location, background objects, resolution, perspective and lighting. All images had adequate image quality to be deemed usable by healthcare professionals.

Alongside images the dataset contains 25 pre-defined pieces of metadata information about the cases which include patient demographics, symptoms, and medical history.
During the labelling process the dermatologists had access to all the captured images alongside all the metadata information.

\section{Experimental Results}
\label{sec:exp}

For model performance we report Top-3 accuracy.
For demonstrating the efficacy of MINT, we show the reduction of inputs required to maintain certain performance metrics. This is reported in 2 ways for both images and metadata questions: (1) average number of inputs versus performance achieved and (2) Histogram of input counts over all cases.

\subsection{Interactive image acquisition}
\label{subsec:int_img_result}
We first report MINT performance on the multi-view interactive image acquisition process.

The trained model, predicting whether to request an additional image or to stop collecting images, achieves a reduction of 27.9\% of images requested, at a difference of $-0.9pp$ in top-3 accuracy.
When additionally predicting the near/far instruction for the next image it achieves reduction of 36.2\% of images requested, at a difference of $-1.1pp$ in top-3 accuracy (see Table \ref{tab:full_results}).

\begin{figure}[t]
\includegraphics[scale=0.6]{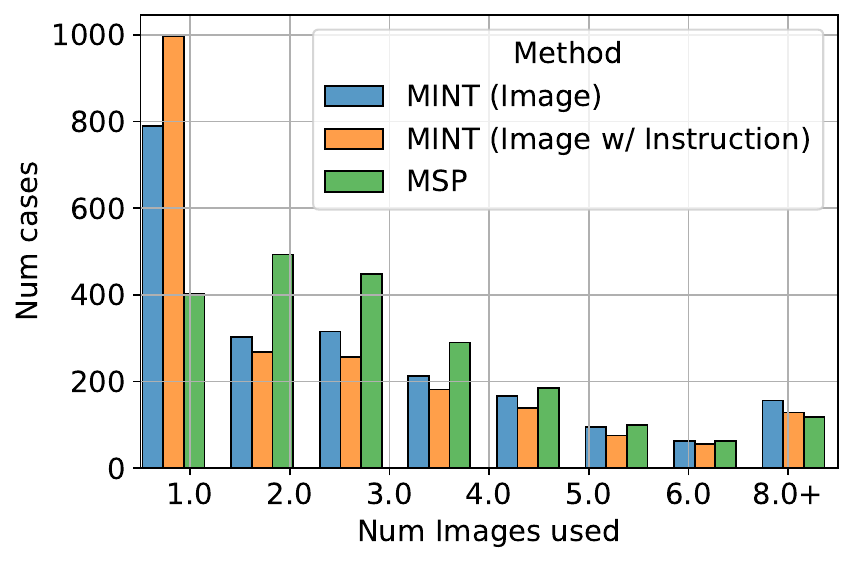}
\caption{Histogram of images used by MINT.}
\label{fig:hist_mint_image_used}
\end{figure}

The histogram in Figure \ref{fig:hist_mint_image_used} shows the number of images requested for cases in the test set.

\begin{table*}[t]
  \centering
  \caption{Results of baselines and our MINT model  using the JS divergence distance metric.}
  \label{tab:full_results}
\begin{tabular}{lccc}
\toprule
Experiment & Top 3 accuracy & Avg. images used & Median num.\\
 & (\%) & (\% reduction) & metadata used\\
\midrule
All image & $56.1$ & $4.37 (-)$ & $0$\\
All image + metadata & $63.1$ & $4.37 (-)$ & $25$\\
\midrule
MSP based ES &$55.1 \pm 0.2$ & $3.4 (21.89)$ & $0$ \\
MINT (Image) & $55.2 \pm 0.2$ & $3.15 (27.86)$ & $0$ \\
MINT (Image w/ instruction) & $55.0 \pm 0.3$ & $2.79 (36.16)$ & $0$\\
\midrule
MINT (Image + metadata) & $61.2 \pm 0.2$ & $3.15 (27.86)$ & $8$ \\
MINT (Image w/ instruction + metadata) & $59.5 \pm 0.3$ & $2.79 (36.16)$ & $4$\\
\bottomrule
\end{tabular}
\end{table*}

\subsection{Interactive tabular metadata acquisition}
\label{sec:int_metadata_result}
In this section, we demonstrate the results of using MINT for interactive acquisition of tabular metadata inputs.
For these experiments, we use a model which combines the image and metadata using a FiLM layer \citep{perez2018film}.
We show MINT results using three different methods of value estimation along with a random and a global baseline.
We report the results on the test set and methods were calibrated (if required) on the validation set.

\subsubsection{Effect of divergence metric for value estimation}
\begin{figure*}[ht]
    \centering
    \includegraphics[width=\textwidth]{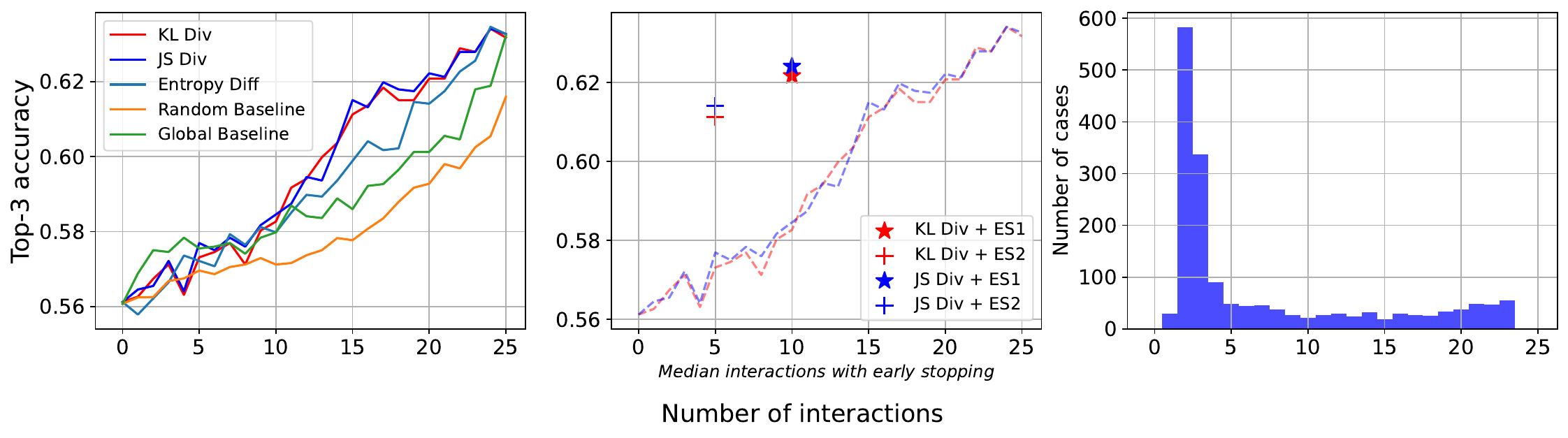}
    \caption{MINT for structured metadata. \textbf{Left:} Top-3 accuracy vs. number of interactions for different divergence metrics and baselines. \textbf{Middle:} Performance and average number of requested pieces of information for two variants of early stopping, using the two best-performing divergence metrics. We include the lines from the left figure for illustration purposes. \textbf{Right:} Histogram of number of interactions using JS divergence + ES1.}
    \label{fig:js_es1_hist_num_int}
\end{figure*}

In this section we present the results for interactive metadata in Fig.~\ref{fig:js_es1_hist_num_int}.
We show MINT with three different divergence metrics for value estimation, i.e. (i) KL divergence, (ii) JS divergence and (iii) absolute difference in predictive entropy as mentioned in Sec.~\ref{sec:mint_method}.

In Fig.~\ref{fig:js_es1_hist_num_int} (left), we show the results when we use the different divergence metrics to re-order the sequence in which metadata inputs are acquired from one case to another. Random baseline simulates the situation where metadata inputs are acquired from different cases in a random order. Global baseline estimates a static sequence of metadata inputs which provides the highest boost on the validation set, and metadata input for all the test cases are acquired in that static order.

Firstly, we observe that the tabular metadata information provides a 7 points boost in Top-3 accuracy to the model.
For all the methods, we observe a steady increase in performance with more interactions/ metadata information.
For the first 10 interactions all the acquisition methods provide a similar boost in performance. Beyond that, we observe that KL divergence and JS divergence provide a significantly greater boost in Top-3 accuracy compared to the random and global baselines.
MINT with absolute difference in entropy as the metric is better than the baselines but worse than the other metrics.
It must be noted that the choice of metric function is a hyper-parameter for MINT and should be chosen separately for different applications.
Also, re-ordering the sequence of metadata inputs using MINT is an unsupervised approach and doesn't require a validation set for calibration. Among all the approaches reported, only global baseline leverages the validation set to estimate the optimal sequence of metadata inputs.

\subsubsection{Effect of Early stopping}
In this section, we investigate the effectiveness of early-stopping we described in Sec.~\ref{subsec:value_estimation}. We present the results in Fig.~\ref{fig:js_es1_hist_num_int} (middle). We limit the results to only the best performing methods: MINT with KL (in red) and JS (in blue) divergences.
For both the methods, we show two different strategies for operating point selection for early stopping, satisfying equations \ref{eq:task1} and \ref{eq:task2} respectively:

\noindent \textbf{Task 1:} Denoted by a $\star$ in the plot. In this setup, we select the value of an early stopping threshold $T$, by defining a tolerance on reduction of Top-3 accuracy. Within a permissible reduction, we choose $T$ to provide the minimum number of median interactions in validation set.

\noindent \textbf{Task 2:} Denoted by a $+$ in the plot. In this setup, we select an operating point (value of threshold $T$) based on a user-defined budget on number of interactions. On the validation set, for a given value of median interaction (here 3), we select a threshold which provides the best Top-3 accuracy.

Firstly, compared to choosing a single fixed number of interactions for every user, early stopping is more flexible, leading to increased performance at the same average number of interactions across users.
Secondly, when attempting to match Top-3 accuracy early stopping leads to significantly fewer interactions required from the user.
This holds for MINT with both KL and JS divergences. Comparing Tasks 1 and 2, we observe that MINT using JS attains slightly better performance than KL as shown in Fig.~\ref{fig:js_es1_hist_num_int} (middle).

In Fig.~\ref{fig:js_es1_hist_num_int} (right), we show the histogram of number of metadata inputs collected over the test set, corresponding to MINT with JS Divergence, calibrated with ES1. We observe a significant number of cases using only 2-3 interactions with MINT in contrast to all 25, reflecting its efficacy in reducing the burden in input acquisition.

\subsubsection{Understanding MINT behavior}

To further understand when and how MINT can save interaction steps, we look at the relationship between the characteristics of a particular case and the number of interactions required when using MINT.

First, we look at the score (summed votes from the dermatologist panel) of the ground truth diagnosis after multi-rater aggregation (for details on this aggregation, see \citealp{liu2020deep}). More difficult cases should result in higher clinician disagreement and therefore a lower score of the diagnosis.
We find a small correlation between case difficulty ($1 - \text{\textit{diagnosis score}}$) and the number of interactions (Spearman's $\rho=0.11$, $p=7.4\times10^{-54}$). For cases for which there is more disagreement between clinicians, MINT asks for more information.

We also look at MINT's behaviour broken down by ground truth class. Since the number of classes is large, we focus on a grouping of classes into risk severity levels: Low, Medium and High (for a full definition, see \citealp{roy2022does}). We find that MINT requests more information for Medium, and even more for High cases (using a 3x1 ANOVA; $F=89.8$, $p=1.5\times10^{-39}$).

We also strive to understand whether MINT asks more relevant questions based on the category of skin condition, such as \textit{Infectious eruption}, \textit{Benign neoplasm}, or \textit{Metabolic eruption}. We compare the frequencies of each metadata question for a given category of skin diseases with the frequency of the question for the population as a whole using a $\chi^2$ test. We find that after Bonferroni correction, 17 out of the 26 categories are asked significantly distinct questions at an alpha level of $0.001$.

\begin{figure}
    \centering
    \includegraphics[width=\columnwidth]{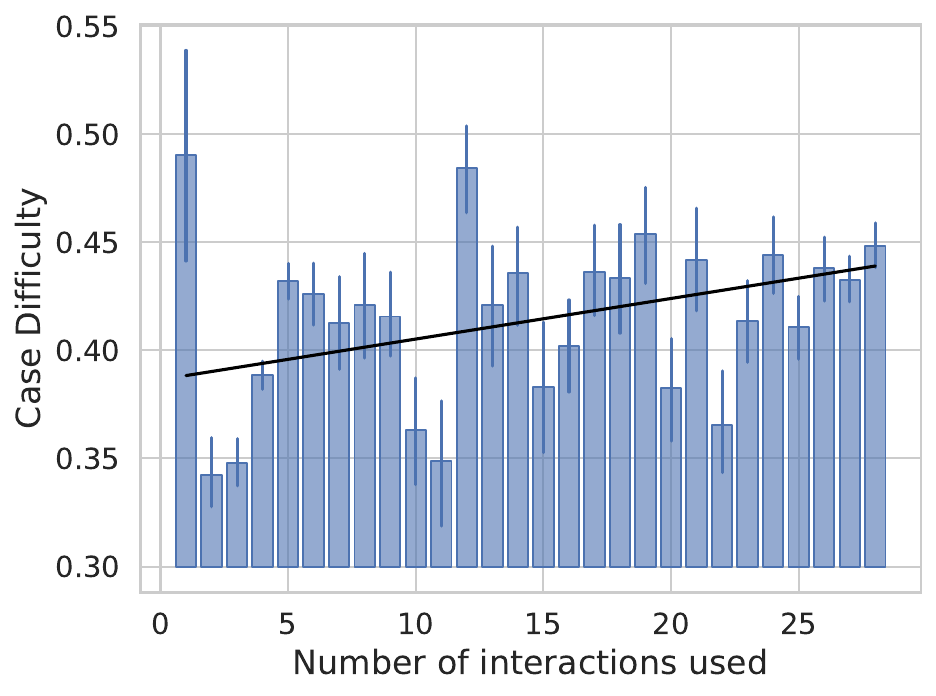}
    \caption{The relationship between the number of interactions requested by MINT, and the difficulty of the case as defined by disagreement between dermatologists. Shown in black is a linear regression (Spearman's $\rho=0.11$, $p=7.4\times10^{-54}$).}
    \label{fig:severity_interactions}
\end{figure}

\subsection{Real-world burden of acquiring information}

Using MINT, we determine how many images and which questions are acquired for each case in our evaluation set. We simulate whether these cases would have successfully completed a real dermatology AI application \citep{dermassist_2022}, using metrics from real users submitting their case to the system.

\begin{figure*}[ht]
    \centering
    \begin{subfigure}[t]{0.33\textwidth}
    \centering
    \includegraphics[width=\columnwidth]{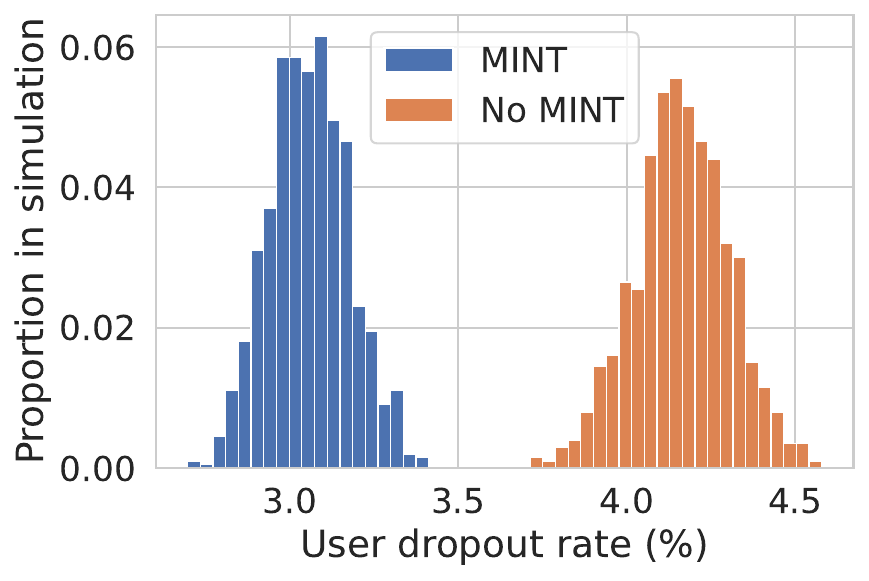}
    \caption{}
    \label{fig:drop_off_results}
    \end{subfigure}
    \hspace{1em}
    \begin{subfigure}[t]{0.33\textwidth}
    \centering
    \includegraphics[width=\columnwidth]{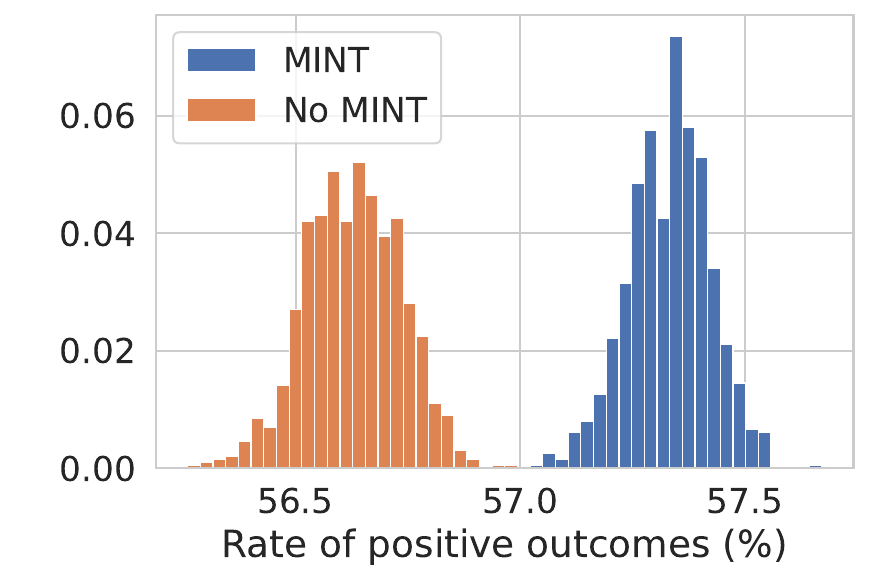}
    \caption{}
    \label{fig:drop_off_outcomes}
    \end{subfigure}
    \hspace{1em}
    \begin{subfigure}[t]{0.25\textwidth}
    \centering
    \includegraphics[width=\columnwidth]{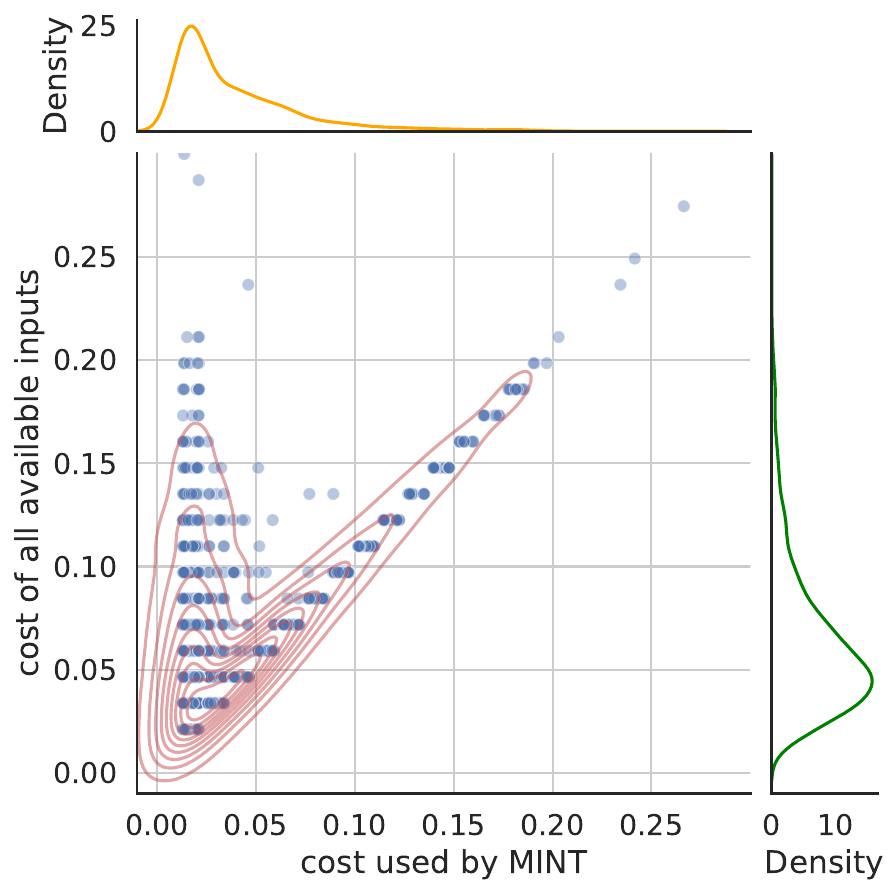}
    \caption{}
    \label{fig:mint_cost_analysis}
    \end{subfigure}
    \caption{Simulated user outcomes. \textbf{(a)} The estimated proportion in users who do not complete the submission flow, with and without MINT. With MINT, we estimate the drop-off rate to be significantly lower. \textbf{(b)} The estimated proportion of users who see a correct result, i.e. who complete the submission flow and for which the diagnosis is in the top 3 predictions. With MINT, significantly more users see correct predictions. \textbf{(c)} Cost analysis using MINT. Contour-lines and density plots were derived using kernel density estimates.}
\end{figure*}

This provides an estimate of the real-world impact of MINT. Of the cases in our evaluation dataset, 4.6\% (3.9–4.4\%) would likely drop off when all information is required. When MINT is used to interactively acquire the information, 3.1\% (2.9– 3.3\%) would likely drop off (see Figure \ref{fig:drop_off_results}.

We also estimate the proportion of users who would be shown a correct result, defined as cases in which the true diagnosis is in the model's top 3 predictions, \textit{and} which did not drop off the simulated submission flow. This analysis, shown in Figure~\ref{fig:drop_off_outcomes}, reveals that even though performance on the validation set reduces slightly with MINT, user outcomes may in fact improve by enabling more users to complete the submission flow and reach the results page.

We further analyse the burden placed on users by assigning each piece of information an empirical cost: its contribution to user drop-off. We compare the cost incurred by MINT and the cost incurred if all data was acquired in Figure \ref{fig:mint_cost_analysis}. The graph highlights that the cost placed on users is bimodal under MINT, echoing the results shown in Figures \ref{fig:hist_mint_image_used} and \ref{fig:js_es1_hist_num_int}.

\section{Discussion}
\label{sec:discussion}

\begin{figure*}[ht]
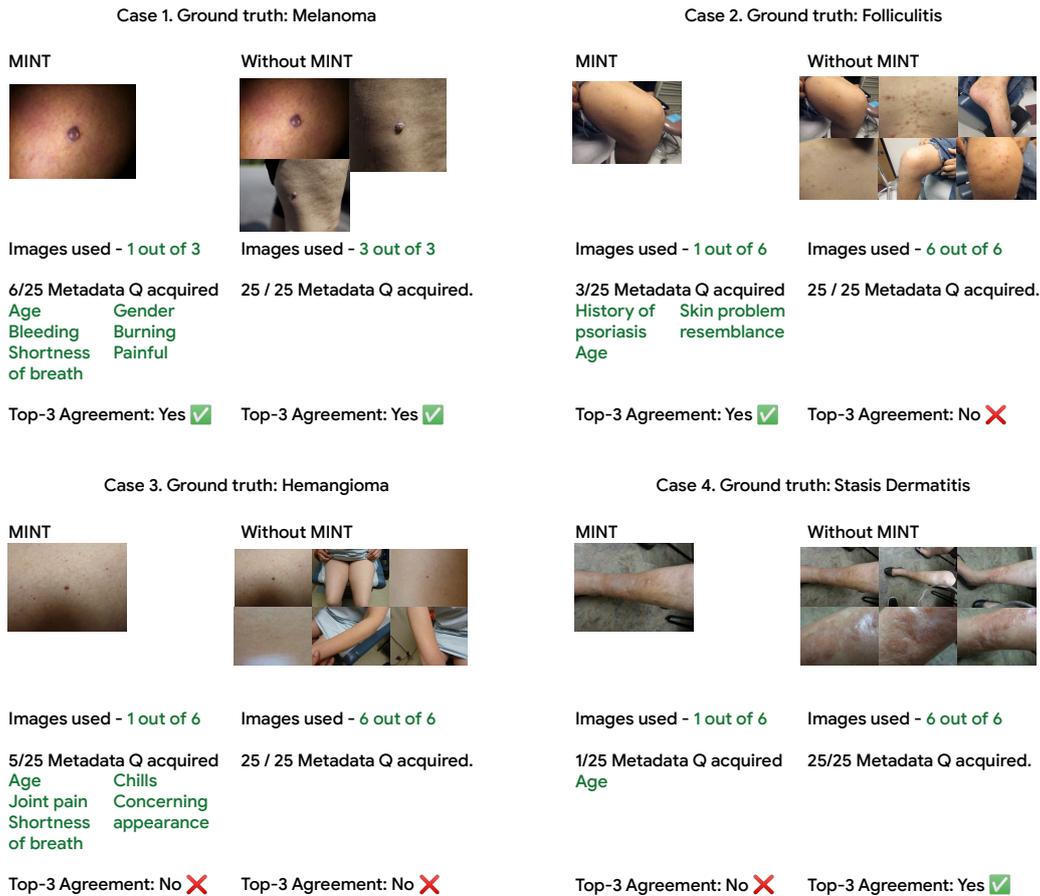

\centering
\begin{subfigure}[t]{0.4\textwidth}
    \adjincludegraphics[width=\textwidth,page=2,trim={{.1\width} {0.05\height}},clip=true]{figures/slide_figures.pdf}
    \end{subfigure}
    \begin{subfigure}[t]{0.4\textwidth}
    \adjincludegraphics[width=\textwidth,page=3,trim={{.1\width} {0.05\height}},clip=true]{figures/slide_figures.pdf}
    \end{subfigure}
    \begin{subfigure}[t]{0.4\textwidth}
    \adjincludegraphics[width=\textwidth,page=4,trim={{.1\width} {0.05\height}},clip=true]{figures/slide_figures.pdf}
    \end{subfigure}
    \begin{subfigure}[t]{0.4\textwidth}
    \adjincludegraphics[width=\textwidth,page=5,trim={{.1\width} {0.05\height}},clip=true]{figures/slide_figures.pdf}
    \end{subfigure}
\caption{Qualitative Analysis of results. Four examples of cases in which processing with and without MINT lead to the same correct result (left), MINT performing better (second), MINT and the model with all inputs both fail (third), and MINT failing while the model with all input succeeds.}
\label{fig:ics_case_study}
\end{figure*}

\subsection{Dynamic feature selection at inference time with equivalent performance}

We propose and evaluate a method for active feature acquisition in the multi-view, multi-modal setting commonly observed in medicine. We demonstrate reduced user burden with approximately equivalent performance by personalising the inputs to each user. We also show that trading off user burden with performance can and should be considered by practitioners. While this is often done before training AI systems, we recommend to instead do so at inference time using MINT.

We show that in a case typical of medical machine learning, we can reduce the number of inputs required from a user by over a third, while maintaining predictive accuracy within a few percentage points. Compared to setting a fixed number of inputs for every user, personalising the inputs each user allows the system to ask for drastically less information in most cases (see Figures \ref{fig:js_es1_hist_num_int} and \ref{fig:hist_mint_image_used}).

\subsection{Simplicity of MINT as a wrapper model}

One important aspect of MINT is its simplicity and generalisability. It is agnostic to the underlying diagnostic model, and has few parameters. This means that MINT can be tuned post-hoc, on a dataset independent of the diagnostic model's training set, and using limited data. We deliberately choose a small model for image value prediction, and use no parameters other than thresholds for metadata value prediction.

This simplicity makes MINT suitable for many other medical applications, and enables practitioners in medical AI to tune MINT parameters separately for a new setting they would like to deploy it in. It also means that MINT is more interpretable than comparable approaches that use reinforcement learning (RL). By directly estimating a piece of information's impact on the predictive distribution, the user can be given the purpose of a new piece of information requested by the system. For example, the user could be told that their age is being asked because it may narrow down the differential diagnosis from 5 to 3 diseases.

\subsection{Easy and difficult cases in MINT}

We believe that one key advantage of MINT is that it is able to stop early when the model does not benefit from further information, i.e. when a case is clearly diagnosable using the information available already. One measure of the ease of diagnosis is clinician agreement. As our data was labelled by multiple clinicians \citep{liu2020deep}, we can use agreement between clinicians as a metric of how unambiguous a case is. Cases for which there was less agreement saw significantly less information requested. So, when a case is clearer, MINT diagnoses faster (using fewer interactions).

We also see MINT asking more questions for high-risk compared to low-risk conditions. While this seems desirable, it is also surprising, since we did not optimise for harm reduction. It may be explained by two factors: that there are more cases of low-risk conditions in the training data (see \ref{tab:severity_distribution}), or that clinician agreement (see above) is lower for high-risk conditions.

\subsection{Qualitative analysis of MINT}
We present some qualitative results using MINT in Fig.~\ref{fig:ics_case_study}.
We contrast the MINT results against our baseline of using all inputs and select 4 random cases for discussion.

Case 1 shows a case where the model makes the correct prediction with and without MINT (1190 out of 2091 cases in the test set). The case has `Melanoma', a high-risk condition as the diagnosis. MINT uses significantly less inputs (1 of 3 images and only 6 out of 25 metadata) information for making the correct prediction.

Case 2 shows an interesting case where the model  makes the correct prediction with MINT but an incorrect one without MINT (50 out of 2091 cases in the test set). The case has `Folliculitis' as its primary diagnosis. MINT uses significantly less inputs (1 of 6 images and only 5 out of 25 metadata) information for making the correct prediction. Using too many inputs caused the model to make the wrong prediction. We believe this is due to noise in certain inputs which MINT can ignore.

Case 3 shows a difficult case where the model makes the incorrect prediction both with and without MINT (721 out of 2091 cases in test set). The patient has `Hemangioma' as its primary diagnosis. Here also MINT uses significantly less inputs (1 out of 6 images, 3 out of 25 metadata). Note from the images that this is indeed a very difficult case as there are freckles present near the mole in near-shot images and the mole is small and hard to see in far-shot images.

Case 4 demonstrates a failure case where the model makes the correct prediction without MINT but an incorrect one when using MINT (138 out of 2091 cases in the test set). The case has `Stasis Dermatitis' as the primary diagnosis. Here also MINT uses significantly less inputs (1 out of 6 images, 1 out of 25 metadata) while being unsuccessful. This may be because the condition is less evident in the image used by MINT.

\subsection{The trade-off between performance and patient burden}

While MINT is flexible and in theory is able to request the upper limit of inputs for a difficult case, there are failure cases. These cases - cases which the diagnostic model predicts correctly with all inputs but where MINT fails to ask all questions - lead to a drop in performance. This means that for practitioners, there is a trade-off between reducing user burden and diagnostic accuracy. Since tuning this trade-off with MINT is relatively simple (choosing a threshold value using a validation set), it is suitable for individualised tuning by practitioners based on use case requirements.

We believe that a key determining factor in this trade off is the rate of users with a completion of the full user flow and seeing correct results. While during development, models may appear to be performing slightly worse with fewer inputs, a lower rate of user dropout due to fewer questions means more users will successfully complete the submission flow and receive accurate results. When simulating patient outcomes using real-world drop-off rates, we actually see an increase in positive user outcomes (see Figure \ref{fig:drop_off_outcomes}). This is despite a small reduction in top 3 accuracy with MINT, and is driven by the increased completion rate of users.

\subsection{User trust and the relevance of questions}

One important issue in medical AI systems is the question of trust. Trust in an application is a multifaceted concept, and can include trust that only relevant data are collected, or trust that the system performs as advertised. Both of these trust domains are affected by the number and types of questions an AI system asks.

By asking fewer questions and ensuring those asked are personalised based on the skin condition, we believe that we are able to improve trust in both domains. Fewer questions and images, alongside more targeted questions reduce the likelihood that a question is seen as intrusive or irrelevant. For example, users looking for information about a small rash may consider a question about cancer history to be of questionable relevance.

MINT enables the AI to personalise questions to be specific to the condition a user is seeking information about, which may engender more trust in the system in the real world.

\subsection{Limitations and future work}

One of the key limitations of MINT is that it is limited by the underlying diagnostic model. This manifests itself in the prediction of relevant questions. If the underlying diagnostic model does not adequately integrate structured questions in its predictions, then predicting which questions are valuable will also be noisy.

The underlying diagnostic model also limits MINT in the types of interactions it can choose. Diagnostic image and metadata models are still widespread in medical AI practice, but may be replaced by large foundation models in the future. In particular, by leveraging natural language, large language models are more flexible both in the types of questions they can ask, and in the breadth of answers that can be provided. While this may make for more efficient and flexible information exchanges, it could also introduce new risks such as asking questions that can be seen as intrusive, as well as providing answers outside of the intended scope. Future work will be needed to understand how best to use MINT on such flexible models.

MINT also lacks long-term temporal modelling over multiple turns of interactions, something which all greedy active feature acquisition methods lack. With larger and more diverse training sets, using reinforcement learning may prove better. These performance benfits of temporal modelling will need to be balanced against the simplicity and robustness of MINT.

\section{Conclusion}
\label{sec:conclusion}

We present MINT, a powerful, practical approach to making medical diagnostic models interactive. We demonstrate its ability to drastically simplify the submission flow for a dermatology system, including the first demonstration of interactive acquisition of multi-image multi-modal data.

We demonstrate the efficacy of this approach using real-world user burden. By simulating user drop off from a dermatology AI system, we estimate that more patients receive a correct diagnosis with less acquired information. This is accomplished with minimal impact on model performance, and highlights the importance of considering user burden in AI system design.

We show that MINT is simple to tune and can be configured to satisfy diverse use cases, allowing practitioners to trade off user burden and model accuracy.

\bibliographystyle{plainnat}
\bibliography{references}

\appendix
\onecolumn

\section{Dataset}

\subsection{Full list of Metadata Questions}

\begin{table}[ht]
    \centering
    \begin{tabular}{l|l}
    \textbf{Question} & \textbf{Options}  \\
    \midrule
    Age     &  \textit{scalar value}    \\
    Gender     & F / M / Unspecified / Prefer not to say / Other \\
    Skin type  & Fitzpatrick skin type 1-7 \\
    Is the appearance concerning? & Yes / No / Unknown\\
    Is it bleeding? & Yes / No / Unknown\\
    Is it burning? & Yes / No / Unknown\\
    Are you experiencing chills? & Yes / No / Unknown\\
    Are you experiencing fatigue? & Yes / No / Unknown\\
    Are you experiencing fever? & Yes / No / Unknown\\
    Are you experiencing joint pain? & Yes / No / Unknown\\
    Are you experiencing joint pain? & Yes / No / Unknown\\
    Are you experiencing mouth sores? & Yes / No / Unknown\\
    Are you experiencing shortness of breath? & Yes / No / Unknown\\
    No symptoms other than what can be seen? & Yes / No / Unknown\\
    Is it itchy? & Yes / No / Unknown\\
    Is it getting darker? & Yes / No / Unknown\\
    Is it getting larger? & Yes / No / Unknown\\
    Is it painful? & Yes / No / Unknown\\
    Do you have a history of eczema? & Yes / No / Unknown\\
    Do you have a history of psoriasis? & Yes / No / Unknown\\
    Do you have a history of melanoma? & Yes / No / Unknown\\
    Do you have a history of skin cancer? & Yes / No / Unknown\\
    Which body part is the skin problem on? & \textit{Select any of 12 body parts} \\
    What best describes your skin issue? & \makecell[tl]{Acne; Growth or mole; Hair loss; \\ Other hair issue; nail issue; Pigment issue} \\
    Duration of problem  & \makecell[tl]{Unknown; Since childhood; One day; Less than one week; \\ One to four weeks; One to three months; Three to twelve months; \\ Over one year; Over five years; Other} \\
    \end{tabular}
    \caption{The full list of metadata questions and the values each can take. For all questions except for age, we represent them as binary vectors with a size equal to the number of options.}
    \label{tab:my_label}
\end{table}

\subsection{Class distribution}

\begin{figure}
    \centering
    \begin{subfigure}{.49\textwidth}
    \centering
    \includegraphics[width=\textwidth,keepaspectratio]{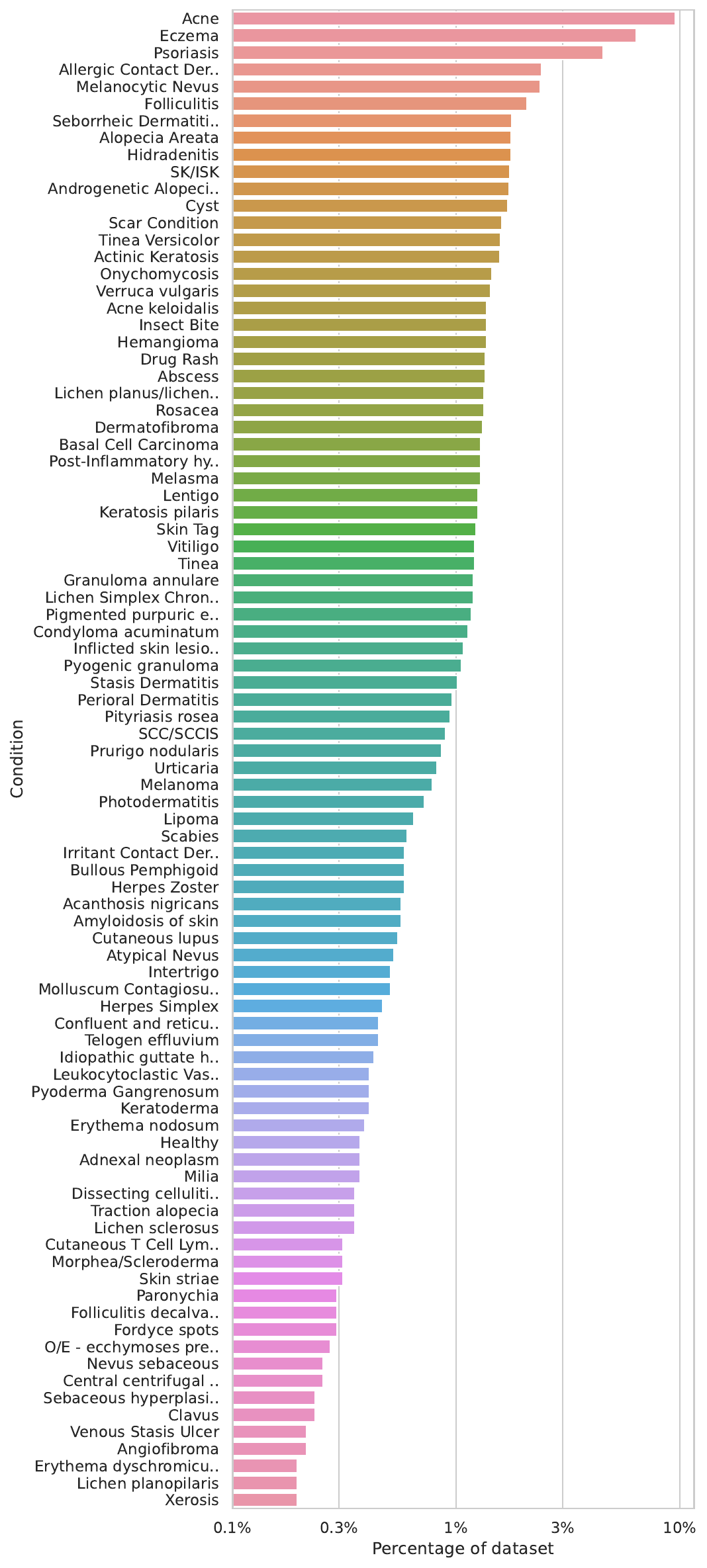}
    \end{subfigure}
    \centering
    \begin{subfigure}{.49\textwidth}
    \centering
    \includegraphics[width=\textwidth,keepaspectratio]{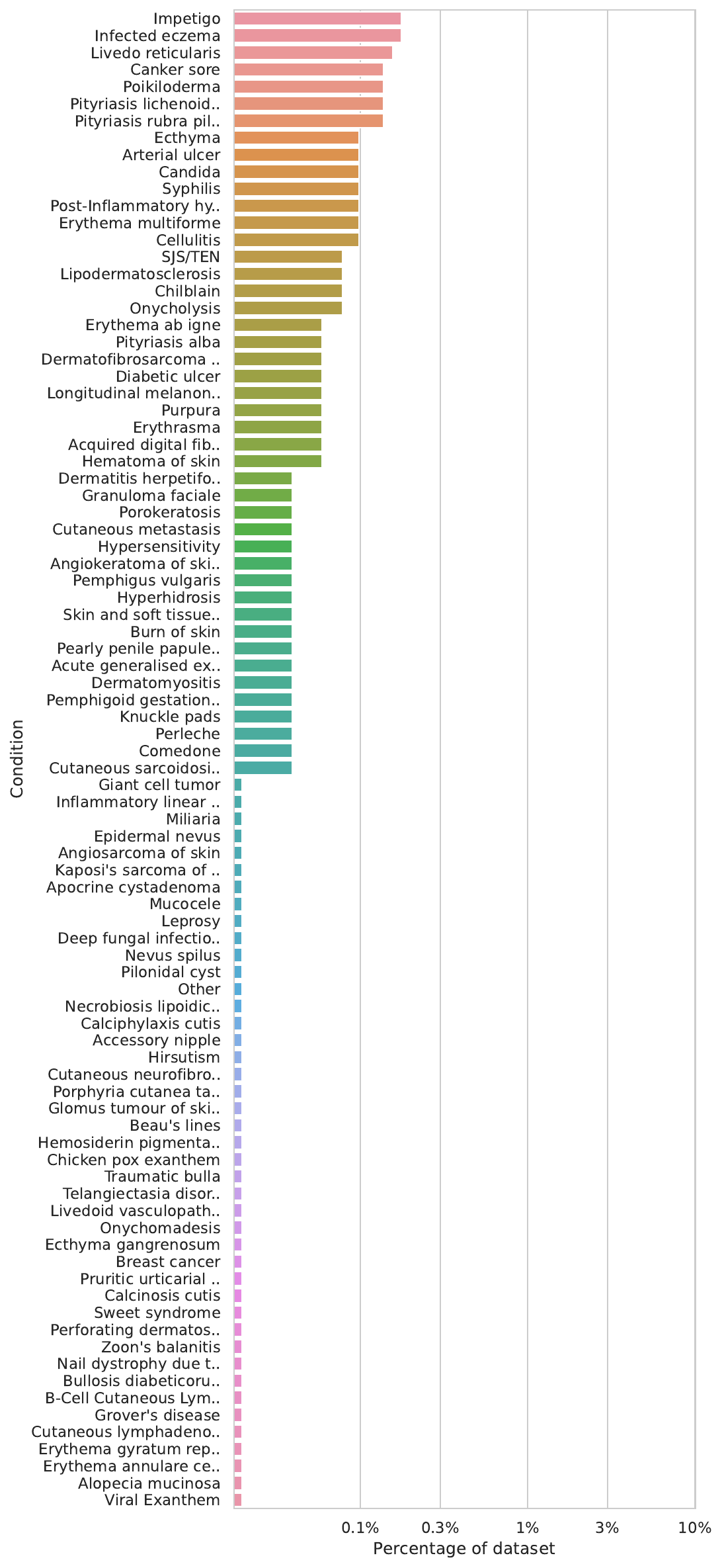}
    \end{subfigure}
    \caption{Percentage of conditions in the dataset}
    \label{fig:condition_histogram}
\end{figure}

For the full distribution of classes in our dataset, see Figure \ref{fig:condition_histogram}.
We also divide conditions into three severity buckets (low, medium and high), the proportions for which are shown in Table \ref{tab:severity_distribution}

\begin{table}[ht]
\caption{Proportion of severity conditions in the dataset.}
\label{tab:severity_distribution}
\centering
\begin{tabular}{ll}
\toprule
\textbf{Severity} & \textbf{Percentage}  \\
\midrule
Low      & 56\%                   \\
Medium   & 36\%                   \\
High     & 8\%                  
\end{tabular}
\end{table}

\section{Training process}

\subsection{Base Model training}
\label{app:model_training}
We resized each image to $448\times448$ pixels for our training. We use data augmentation for training. This includes: random horizontal and vertical flips, random variations of brightness (max intensity = $0.1$), contrast (intensity = [$0.8-1.2$]), saturation (intensity = [$0.8-1.2$]) and hue (max intensity = $0.02$), random Gaussian blurring using standard deviation between $0.01$ and $7.0$ and random rotations between $-150^{\circ}$ and $150^{\circ}$.

In order to ensure the model is robust to variable numbers of input, image embeddings are averaged, producing a single average embedding for each case. Additionally, we train the model with random metadata masking.

We use a Stochastic Gradient Descent optimiser with momentum and exponentially decaying learning rate for training. Each model is trained for $10\,000$ steps. Convergence of all the models were ensured on the validation set. We use a batch-size of $16$ cases for training.
We use the validation set to set the hyper-parameters ( initial learning rate, decay factor, momentum) and to select the best checkpoints for all the models. Checkpoint selection was based on Top-3 accuracy.

\end{document}